# Probing of Excitonic Transitions in All Inorganic Perovskite Quantum Dots ($CsPbX_3$: X= Cl, Br, I) by Magnetic Circular Dichroism Spectroscopy

Prasenjit Mandal [b] and Ranjani Viswanatha*[, a, b]

[a] *International Centre for Material Science (ICMS), JNCASR, Bangalore*

[b]*New Chemistry Unit (NCU), JNCASR, Bangalore*

*Corresponding author, email: rv@jncasr.ac.in

**Abstract**

Higher-order electronic transitions of all inorganic lead halide perovskite quantum dots (QDs) ($CsPbX_3$: X = Cl, Br, I) are hardly detected by traditional spectroscopic techniques due to the condensed electronic level of QDs. This work assigned the various excitonic transitions in undoped $CsPbX_3$ QDs through a high-resolution absorption spectroscopic technique - magnetic circular dichroism (MCD). Moreover, we investigated the nature of these transitions through sensitive Zeeman responses to the electronic states, likely involving the interaction of spins with the applied external magnetic field. The study unveiled multiple excitonic transitions in the QDs, showing unusual temperature dependence and shedding light on the role of spin-orbit interactions in the presence of a magnetic field. This study offers a comprehensive insight into excitonic transitions, and the potential manipulation of their spin through thermal means holds the promise of advancing electronic and photonic devices based on these materials.

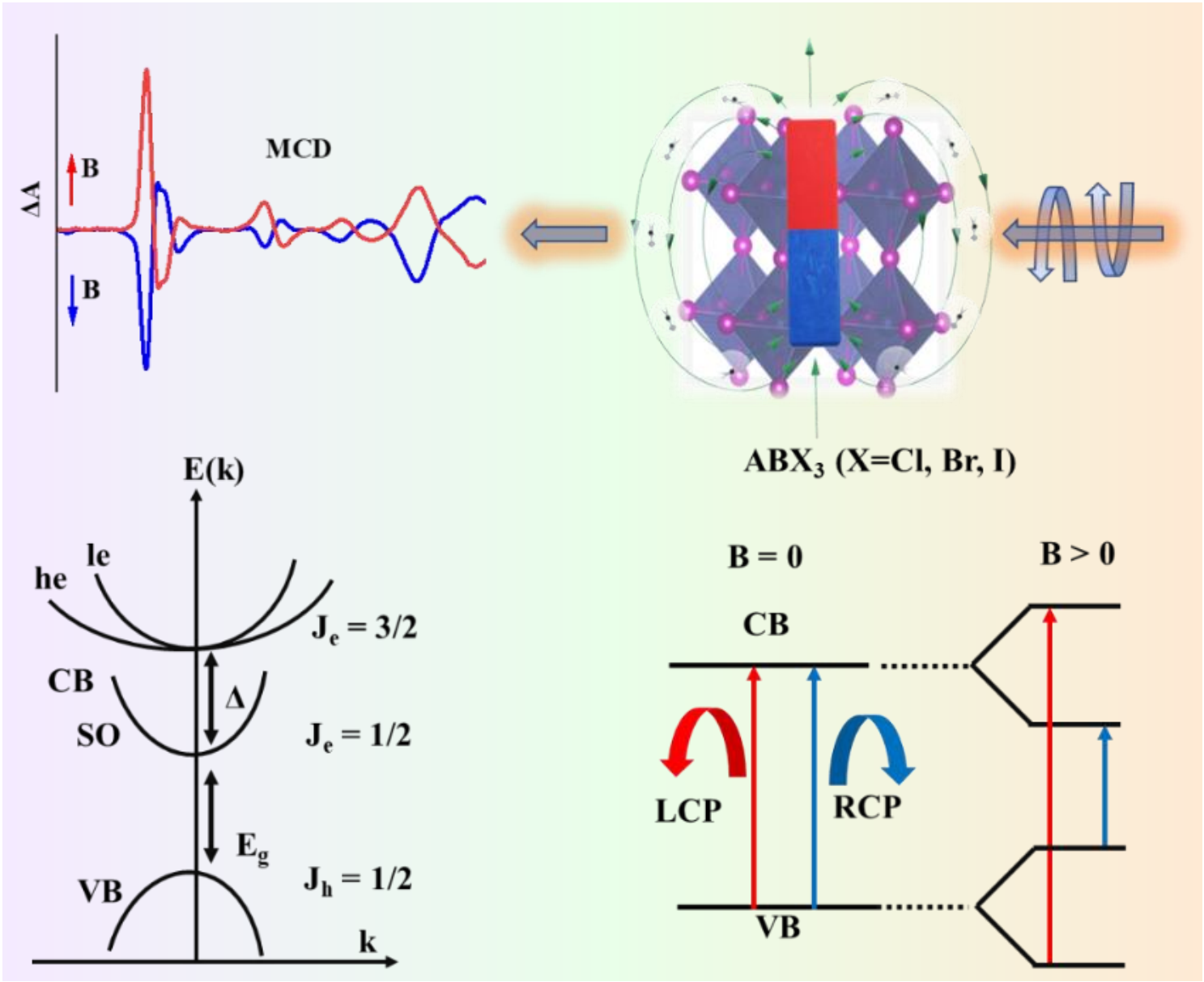

**Figure** TOC of the paper

## Introduction

Perovskite halides have been the subject of increasing interest for their potential applications in spintronics, spin-based photovoltaics, optoelectronics, and light-emitting devices (LED).[1] The key advantages of perovskite halides in this field are their high photoluminescence quantum yield (PLQY), bandgap tunability over the entire UV-Vis region, solution processability, high carrier mobility, high absorption coefficient, strong spin-orbit coupling and high spin-relaxation time, etc.[2-7] Some of these advantages were recently used to show a perovskite-based single photon emitter, the production of white light, a spin-LED that works at room temperature, photodetector, photocatalytic $CO_2$ reduction, and a potential material for read-out of spins.[8-14]

Perovskite halide nanocrystals (NCs) of $CsPbX_3$ are typically cubic in shape.[15-16] However, the shape of these NCs can range from cuboidal to spherical to nanoplatelet to hexapods to even more complex structures depending on the synthesis conditions and the method used to prepare the materials.[17-21] Due to shape and size modulation, these novel materials exhibit a wide range of optical properties. Such unique properties result from the electron and hole's three-dimensional confinement within the QDs. This happens when the size of the NCs is similar to or smaller than the semiconductor's exciton Bohr radius. Discrete excitonic states dominate the photo-physics behind these strongly confined NCs. These states contribute to well-defined excitonic peaks in the absorption spectrum and many other properties. Bawendi et al. reported the evolution of ten such absorption features in CdSe QDs.[22] Recently, Akkerman et al. reported that spheroidal $CsPbX_3$ quantum dots exhibit up to four well-resolved excitonic transitions.[23-24] In contrast, cubic $CsPbX_3$ NCs do not show clear second, third, or higher-order absorption transitions in the absorption spectra. This is mostly because cubic symmetry introduces a perturbation. This perturbation couples the spherical state of different angular moments and causes mixing between and splitting of the higher-order absorption states. As a result, it leads to the smoothing out of the absorption spectra seen in experiments. Therefore, traditional optical-absorption spectroscopy barely detects all the transitions.

Circular dichroism (CD) is widely used in chiral molecules for validating the materials' chirality and excitonic transitions. However, all inorganic perovskite halides ($CsPbX_3$) lack inherent chirality because of their centrosymmetric structure. As a result, research into the detailed investigation of their optical transitions and electronic structures is limited. To overcome these limitations, magnetic circular dichroism (MCD) spectroscopy, which is based on the distinction between the absorption of right circularly polarised (RCP) light ($A_R$) and left circularly polarised (LCP) light ($A_L$) in the presence of a magnetic field along the incident light, can be utilized to observe condensed electronic energy levels in cubic NCs. MCD can be expressed as $\Delta A = A_R - A_L$. Basically, in the presence of a magnetic field, the electronic states involved in optical transitions no longer remain degenerate. Instead, they split into additional energy levels due to various magnetic transitions, including the Zeeman effect. This splitting gives rise to spectral signals in MCD in accordance with the selection rules.

Hamiltonian for the Zeeman splitting is given by $H = -\mu \cdot B$ where μ is the magnetic moment of the sample and B is the magnetic field. This describes the interaction energy between the magnetic moment and the magnetic field. MCD spectroscopy is widely used to study the detailed information about the electronic states of NCs, including characteristics related to the $g$ factor of electrons, holes, and excitons in QDs, the dopant-carrier *sp-d* exchange effect of diluted magnetic semiconductor NCs (DMSs), ligand field transition, magnetic exchange interaction etc.[25-31] The Faraday and optical Kerr rotations are historically well-known examples in the field of perovskite halides. Lakhwani et al. successfully demonstrated the Faraday rotation in nonchiral perovskite single crystals of methyl ammonium lead-halides.[32-33] Huang et al. observed the magneto-optical Kerr effect (MOKE) in chiral hybrid organic–inorganic perovskites (chiral-HOIPs).[34] However, magneto-optical studies of lead-based perovskites NCs' using MCD spectroscopy are not much explored. Recently, Jacoby et al. employed MCD spectroscopy to measure $g_{ex}$ in vapor-deposited $CsPbBr_3$ thin films.[35] Canneson et al. used time-resolved and polarization-resolved spectroscopy in the high magnetic field to reveal the recombination dynamics in $CsPbBr_3$ NCs at cryogenic temperature.[36] Walsh et al. studied the spin polarised luminescence in $CsEuCl_3$ NCs in the presence of a magnetic field.[37] Therefore, a comprehensive and methodical investigation of the optical transitions and Zeeman interactions in achiral $CsPbX_3$ NCs in the presence of a magnetic field and circularly polarised light can shed light to the various electronic states participating in the transitions.

In this work, for the first time, MCD spectroscopy was used to detect up to six excitonic transitions in the cubic all-inorganic perovskite lead halide $CsPbX_3$ NCs. These transitions are assigned based on spectroscopic criteria. This technique was chosen because of its proven ability to probe the condensed energy levels due to its' polarizability dependent specific sign and selection rules. For these undoped NCs, we noticed an unusual behaviour of temperature dependence on the MCD spectra, and a systemic investigation was done on Zeeman splitting energy for the assigned transitions. It is concluded that heavier lead and halogen-induced spin-orbit interaction plays a significant role in this temperature-dependent behaviour.

## Experimental methods

### Chemicals

In the experiments, the following chemicals were utilised: $Cs_2CO_3$ (99.9%), Oleyl amine (OAm) (technical grade 70%), Oleic acid (OAc) (technical grade 90%), 1-Octadecene (ODE) (technical grade 90%), $PbCl_2$ (99.9%), $PbBr_2$ (98%) $PbI_2$ (98%), n-hexane (>97.0%), methyl acetate, anhydrous (99.5%), and trioctylphosphine (TOP). All these chemicals were purchased from Sigma Aldrich. Hexane (AR grade) was bought from Thermo Fischer Scientific India Pvt. Ltd. None of the chemicals were purified before use.

### Methods

### Preparation of cesium oleate (Cs-oleate)

20 mL of ODE and 1.5 mL of OAc were mixed with 400 mg (1.23 mmol) of $Cs_2CO_3$ in a 50 mL three-necked round bottom flask. The flask was degassed for 1.25 hours while continuously stirring at 120 °C. Then, argon was purged into the degassed solution for 15 to 20 minutes at 120 °C. The temperature was then gradually raised to 140 °C in an argon environment and maintained there until all of the $Cs_2CO_3$ had reacted with the oleic acid to produce Cs-oleate. The solution was then brought to room temperature and put into vials filled with argon for later use. To get a clean solution before hot injection, the solution was heated to dissolve the precipitate.

**Synthesis of $CsPbX_3$ (X= Cl, Br, I) NCs**

The hot injection technique described by Protesescu et al. was used to make a series of all-inorganic perovskite nanocrystals of $CsPbX_3$ by changing the composition of the halides.[16] In a typical synthesis, 0.188 mmol of Pb-halide salt, 5 mL of ODE, 1 mL of OAc, and 1 mL of OAm were poured into a three-neck round bottom flask. At 120°C, the contents were stirred and degassed for 1 hour to get rid of all the gases. After that, a specific amount of TOP was employed, and the system was again degassed for 15 minutes. Then the temperature is gradually increased to 170°C under an argon atmosphere. At 170°C, 0.4 mL of pre-heated Cs-oleate was swiftly injected into the flask, and the reaction was quickly quenched in an ice bath. A similar synthesis procedure was followed to obtain mixed halide NCs. For the synthesis of $CsPbI_3$, 0.2 mmol of $PbI_2$ along with 4 mL of ODE, 0.5 mL of OAc, and 0.5 mL of OAm were loaded into a three-neck round bottom flask. At 110°C, the contents were stirred and degassed for 1 hour to eliminate all the gases and then the temperature was increased to 150°C to dissolve the Pb-salt. Then the solution temperature was decreased to stabilize at 60°C. 0.2 mL of pre-heated Cs-oleate was swiftly injected into the flask and stirred for 10 min to get the $CsPbI_3$ NCs. The details amount of the precursors are listed below in Table T1.

---

**Table T1** Amount of precursors used for the synthesis of pure and mixed halide NCs.

| Sample Name | $PbCl_2$ (mg) | $PbBr_2$ (mg) | $PbI_2$ (mg) | TOP (mL) |
|---|---|---|---|---|
| P1- $CsPbCl_3$ | 52.28 | - | - | 1.0 |
| P2- $CsPb(Cl_{0.8}Br_{0.2})_3$ | 41.82 | 13.79 | - | 1.0 |
| P3- $CsPb(Cl_{0.6}Br_{0.4})_3$ | 31.36 | 27.59 | - | 1.0 |
| P4- $CsPb(Cl_{0.4}Br_{0.6})_3$ | 20.91 | 41.39 | - | 0.5 |
| P5- $CsPb(Cl_{0.2}Br_{0.8})_3$ | 10.45 | 55.19 | - | 0.5 |
| P6- $CsPbBr_3$ | - | 69 | - | - |
| P7- $CsPb(Br_{0.8}I_{0.2})_3$ | - | 55.19 | 17.33 | - |
| P8- $CsPb(Br_{0.6}I_{0.4})_3$ | - | 41.39 | 34.66 | - |
| P9- $CsPb(Br_{0.5}I_{0.5})_3$ | - | 34.49 | 43.33 | - |
| P10- $CsPbI_3$ | - | - | 92.2 | - |

---

**Purification of synthesized NCs**

After quenching in an ice bath (as detailed in the preceding section), the contents of the flask were centrifuged at 5000 rpm for 10 minutes. The supernatant was removed after centrifugation, and the particles were re-dispersed in 4 mL hexane and refrigerated for 24 hours. Then the supernatant was washed with methyl acetate, and the precipitate was dissolved in hexane. This solution was kept under refrigeration for further characterization.

**Characterization and spectroscopic studies**

Various techniques were used to characterise and analyse the synthesized nanocrystals. Using a JEOL JEM-2100 Plus transmission electron microscope with an accelerating voltage of 200 kV and the bright field imaging technique, microscopic imaging was performed. For sample preparation, a drop of purified NCs dissolved in hexane was placed on a carbon-coated Cu grid. The solvent was left to evaporate, leaving behind the NCs for imaging. The NCs' crystal structure was identified using XRD, which was measured using a Bruker D8 advance diffractometer and a Rigaku Smartlab using Cu-$K_\alpha$ radiation. Using an Agilent 8453 UV-visible spectrometer, UV-visible absorption spectra of perovskite NCs dissolved in hexane were recorded. The FLSP920 spectrometer from Edinburgh Instruments was used to gather steady-state PL spectra with a 450 W xenon lamp as the source, while the PL lifetime measurements were carried out in the same instrument using EPL-405 pulsed diode laser and EPL-340 diode laser as the excitation sources (λex = 405 nm and λex =340 nm). MCD measurements were carried out on a customized setup where we coupled a CD spectrometer with a magnet. The wavelength-tunable light source was based on a 450 W water-cooled Xe lamp and a double prism polarising monochromator (JASCO J-1500-450). The sample was drop cast on a spectrosil B quartz slide and placed between two poles of the spectromag PT with a 7T split pair solenoid magnet (Oxford Instruments) in Faraday geometry where a cryostat is placed for temperature-dependent measurement. The polarized light passes through the sample, and a photomultiplier detects the transmitted light.

## Results and Discussions

To synthesize lead halide perovskite NCs of $CsPbX_3$, we used the standard colloidal hot injection method, and the detailed synthesis procedure is discussed in the experimental section. The basic characterization of purified perovskite NCs are depicted in Figure 1. Spectroscopic characterization has been carried out by UV-visible and photoluminescence (PL) spectroscopy. Figure 1a depicts the UV-visible and PL spectra of perovskite NCs dissolved in hexane at room temperature. Both absorption and emission spectra red shifted towards higher wavelength with the change in the perovskite composition from pure $CsPbCl_3$ (P1) to mixed halide to pure $CsPbI_3$ (P10). It is important to note that in the absorption spectra except the band edge transition higher order transitions are not clearly visible. Further the samples were characterized through the time resolved decay plots obtained at band edge of typical NCs are presented in Figure 1b. All sample emission lifetimes are fitted with a multiexponential decay law based on equation 1.

**Equation 1** $$I(t) = \sum_{i=1}^{n} a_i e^{-t/\tau_i}$$

where $I(t)$ is the time-dependent fluorescence intensity, $\tau_i$ is the decay time of i th component, and $a_i$ represents the corresponding amplitude. The calculated amplitude-average lifetime of the samples increased from 2.5 ns in P1 to 36 ns in P10. Further, structural characterization has been carried out by X-ray diffraction (XRD) patterns and transmission electron microscopy (TEM). Figure 1c displays XRD pattern data for these NCs. These data are consistent with cubic $CsPbCl_3$ and cubic $CsPbBr_3$ ICSD database for the pure Cl (P1) and pure Br (P6) samples and suggest the absence of crystalline impurity phases. For the mixed halide samples, a slight peak shift is seen because of the compositional variance. Figure 2 (a-i) depicts TEM images of the NCs from P1 to P9, demonstrating the formation of cubic nanoparticles that are uniformly distributed. We conducted an analysis of more than 250 nanoparticles to determine their average size, as illustrated in Figure 3 (a-i). The average size of nanoparticles falls within the range of 10-12 nm.

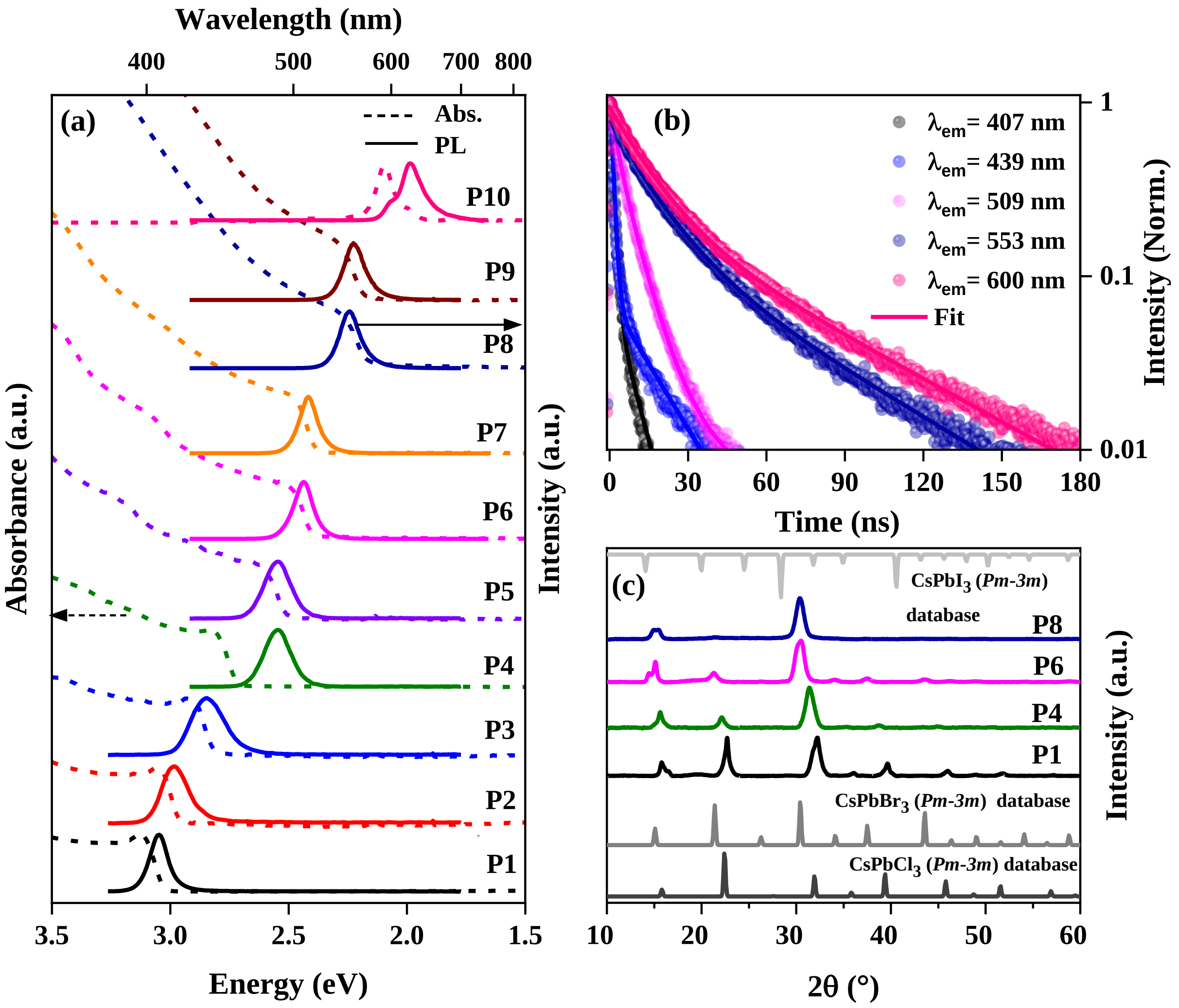

**Figure 1** Spectroscopic and structural characterization of the synthesized perovskite NCs. (a) Abs./PL spectra of the colloidal solution of the nanocrystals dispersed in hexane, (b) Lifetime decay plot obtained using 340 nm and 405 nm laser as a excitation source along with the fit, (c) XRD crystal structures along with the ICSD database.

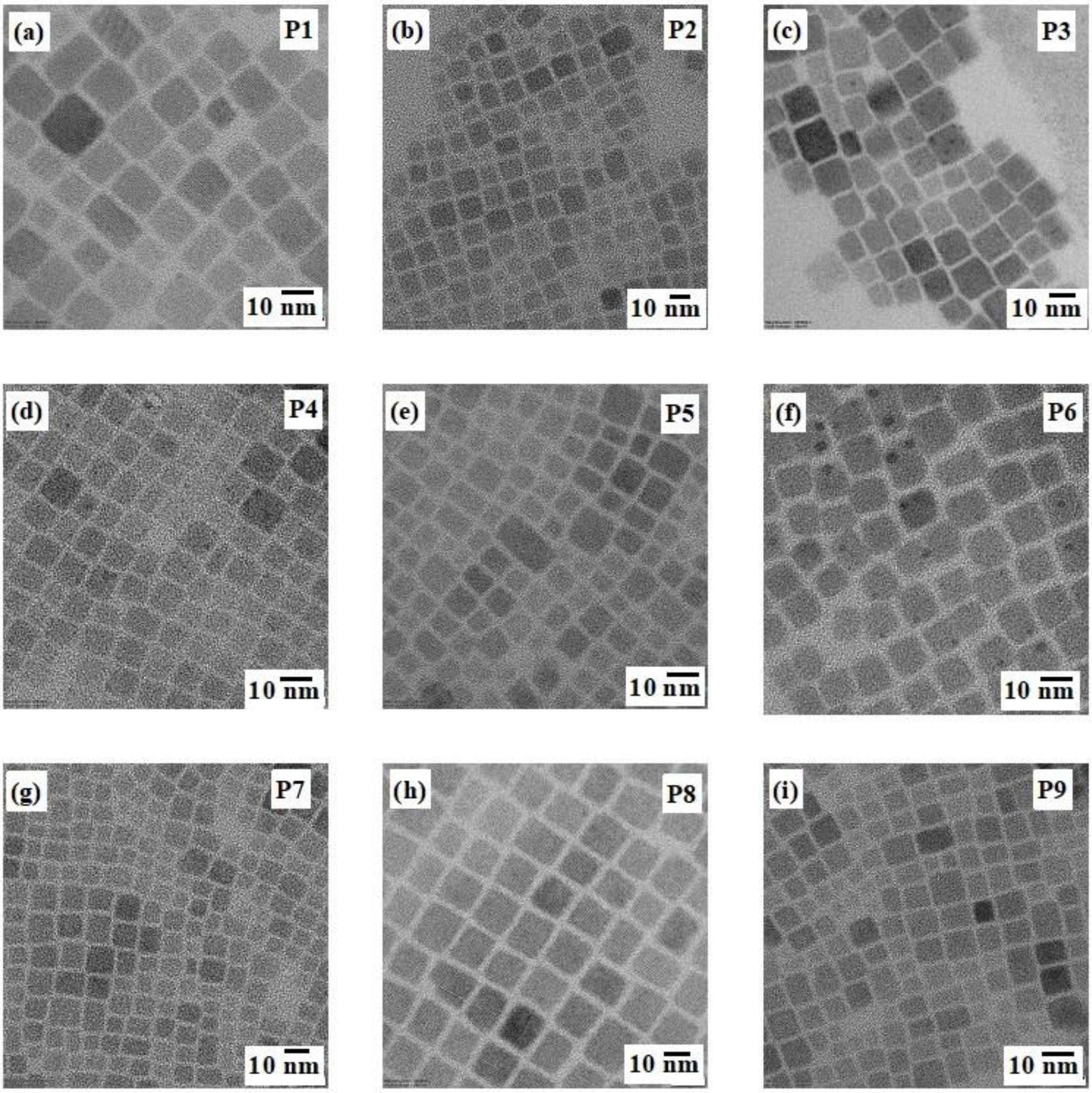

**Figure 2** (a-i) TEM images showing the formation of monodisperse cubic NCs all over the samples from P1 to P9.

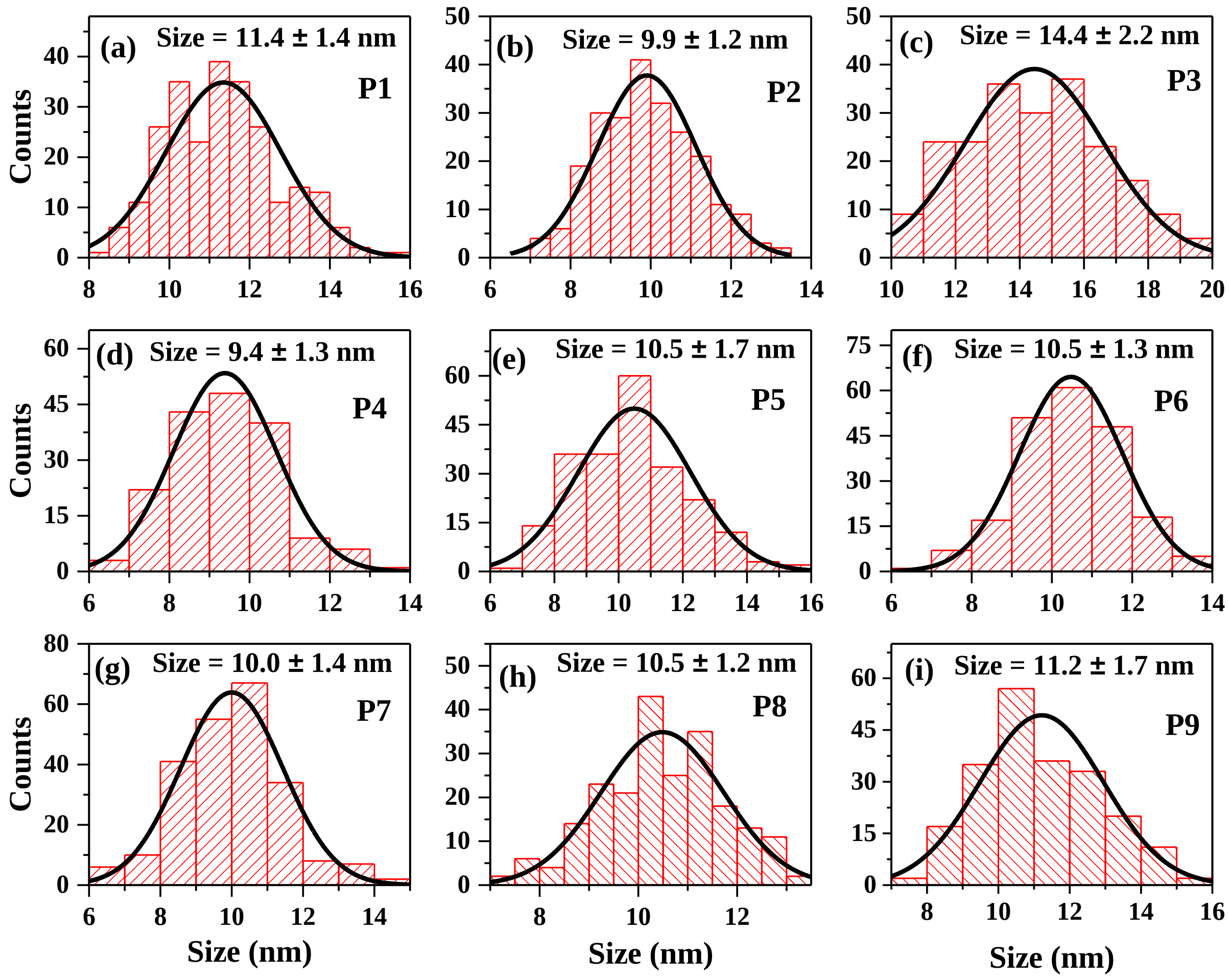

**Figure 3** Histogram of size analysis of the NCs. The mean value of gaussian distributions is used to construct the sizing curve. a) P1 (Mean size 11.4 ± 1.4 nm), b) P2 (Mean size 9.9 ± 1.2 nm), c) P3 (Mean size 14.4 ± 2.2 nm), d) P4 (Mean size 9.4 ± 1.3 nm), e) P5 (Mean size 10.5± 1.7 nm), f) P6 (Mean size 10.5 ± 1.3 nm), g) P7 (Mean size 10.0 ± 1.4 nm), h) P8 (Mean size 10.5 ± 1.2 nm), and i) P9 (Mean size 11.2 ± 1.7 nm).

---

Once we confirm the formation of phase pure NCs using the above-mentioned technique, we move on to studying the magnetic-optical properties of the sample. For that, on quartz (spectrosil-B) substrates, thin films of these NCs were made by drop-casting nanocrystal dispersions. The sample was put in an external magnetic field in the Faraday geometry, and the differential absorption of right and left circularly polarised light was measured to determine magnetic circular dichroism. This setup allows us to concurrently measure the absorption and MCD spectra. Figure 4a depicts the absorbance and associated 3- Tesla MCD signal of P1 at temperature of 1.5 K with the change in the magnetic field direction. The MCD spectrum shown in Figure 4b has a distinct first derivative line shape, and the MCD signal passes zero at the first absorption transition peak. We have also noticed that when the magnetic field direction is inverted, the spectra are simply flipped. This suggests that the CD signals are caused by the magnetic field aligning with the incident light, and that the NCs do not possess inherent chirality. Additionally, the intensity of MCD exhibits a linear relationship with the applied magnetic field's strength, as depicted in Figure 4c. This demonstrates the diamagnetic nature

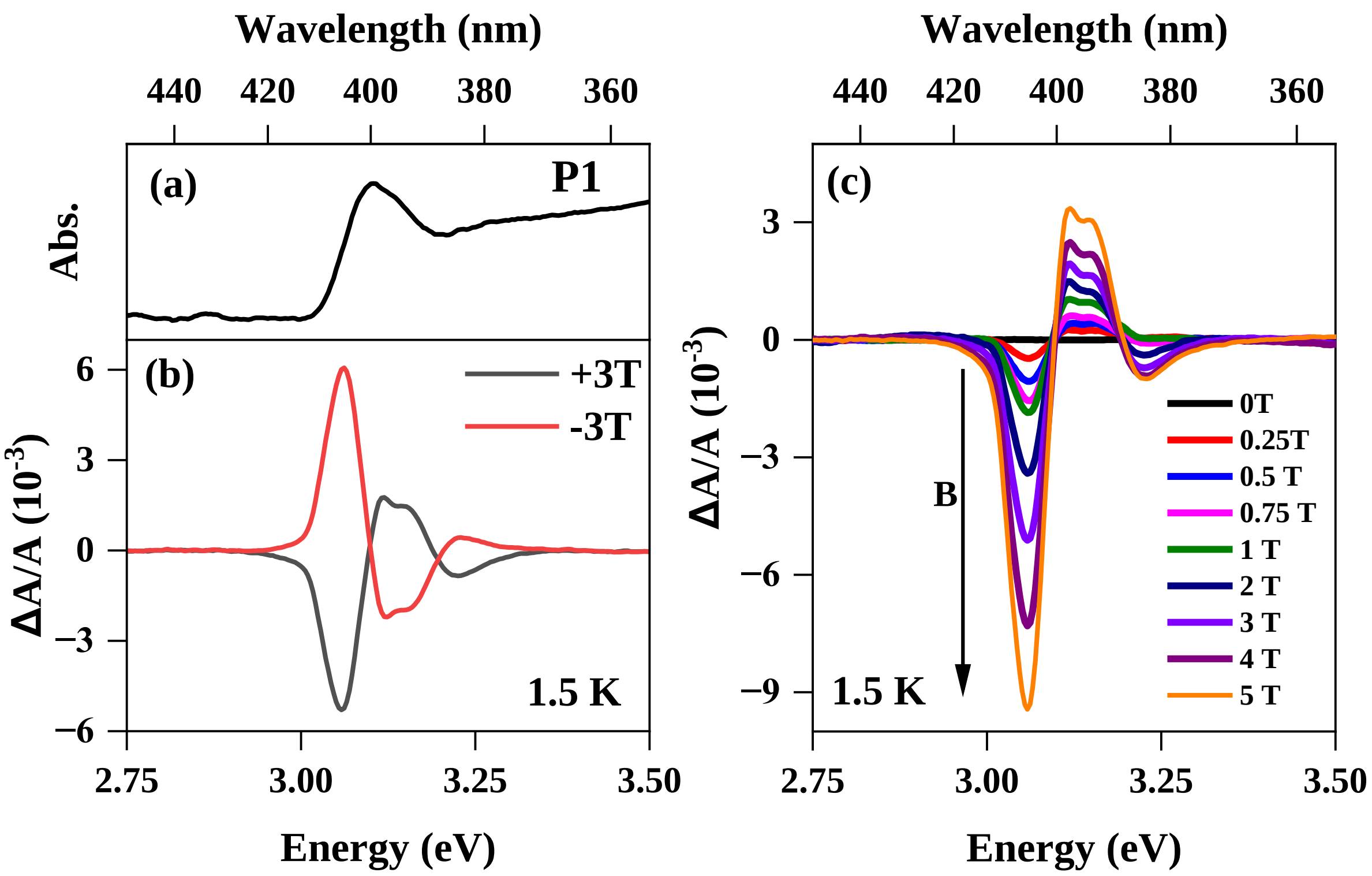

**Figure 4** (a) Absorption spectra of P1 at 1.5 K, (b) MCD spectra of P1 at 1.5 K and 3T magnetic field with the change in the magnetic field direction. (c) MCD spectra of P1 at 1.5 K with various magnetic fields from 0 to 5T.

of perovskite NCs, which is analogous to conventional II-VI semiconductor QDs.[31] Therefore, it is expected that the temperature's influence would be negligible, except for the variations in linewidth and peak shift caused by lattice thermal motion. It is well-documented that temperature can influence the spin interaction in II-VI semiconductor QDs when they are doped with paramagnetic ions.[38-40] However, upon conducting variable-temperature MCD measurements on undoped perovskite NCs, we observed a temperature-dependent behaviour, as illustrated in Figure 5. This behaviour is unexpected due to lack of temperature dependent term in the Hamiltonian. Figure 5 illustrates the typical MCD spectra of P2 and P4 at temperatures of 1.5 K, 3 K, 5 K, 10 K, and 20 K in a 3T magnetic field. The MCD spectrum suggests that P2 (Br 20%) (Figure 5 a) is independent of temperature, whereas P4 (Br 60 %) (Figure 5b) is extremely temperature dependent. This suggests that increasing heavier halide content plays a significant role in describing the magnetic interaction at the band edge.

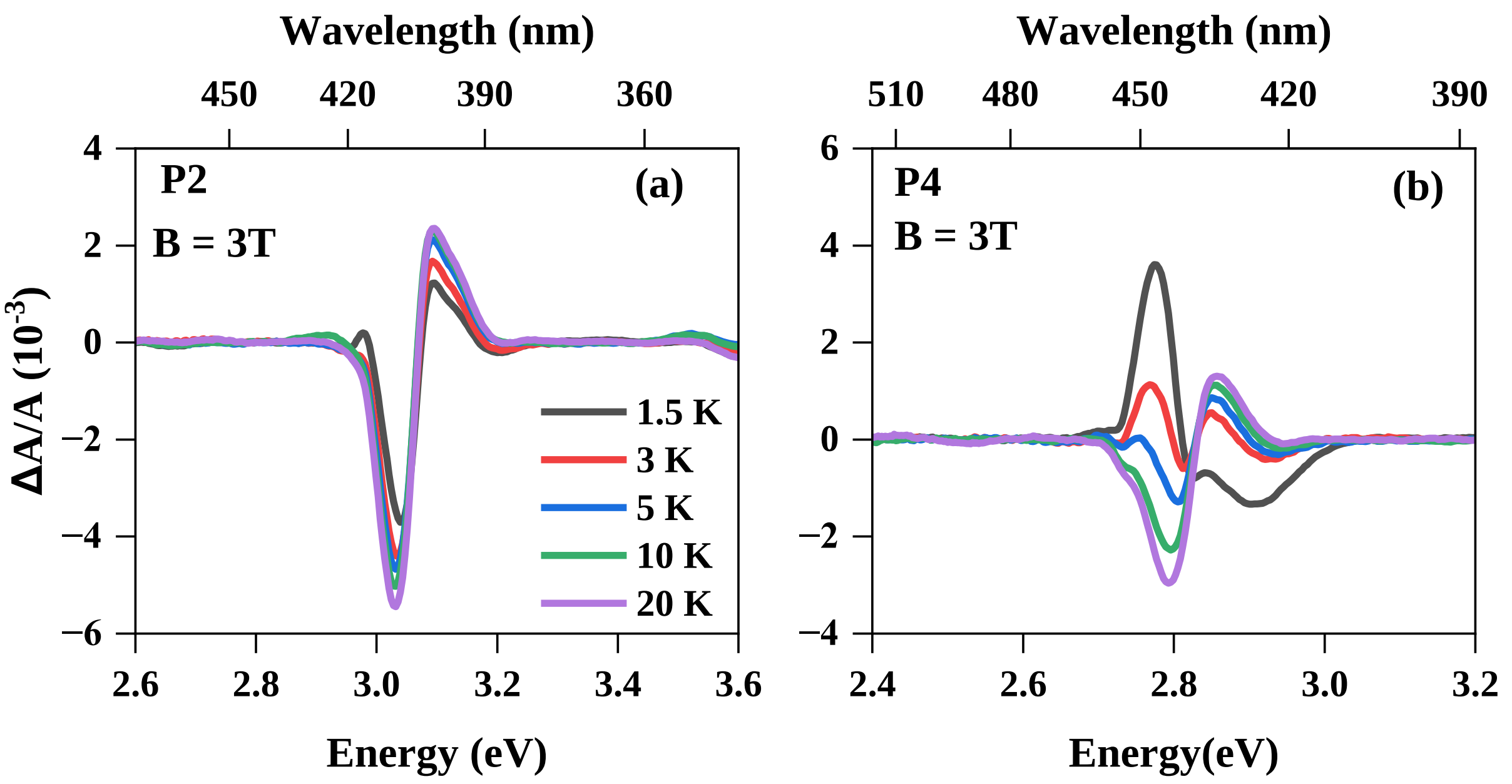

**Figure 5** MCD spectra of (a) P2 - $CsPb(Cl_{0.8}Br_{0.2})_3$ and (b) P4- $CsPb(Cl_{0.4}Br_{0.6})_3$ at a fixed magnetic field of B = 3T with various temperatures.

Thus, a systemic investigation of electronic structure is required. The Pb and X atomic orbitals make up the majority of the electronic band structure that is located close to the band gap. For cubic band structure without spin-orbit coupling (SOC) the edge of the valence band is made up of X *p*-orbitals combined with Pb *s*-orbital contribution with $R_1^+$ (*s*-like) symmetry Bloch function. The edge of the conduction band is made up of Pb orbitals with $R_4^-$ (*p*-like) symmetry wave function. Thus, arrangement of energy levels in the perovskite band structure is opposite to that observed in conventional semiconductors.[41-42] The presence of heavy metal lead (Pb) causes significant spin-orbit coupling (SOC), which split the triply degenerate conduction band minima at $R_4^-$ into doublet split off level (in spin) $R_{6v}^{(2)}$ $(J = \frac{1}{2})$ and *p*-like quartet level (in spin) $R_{8v}^{(4)}$ $(J = 3/2)$. However, with SOC, there are no significant changes in the VB level. In this case, the state is represented as $R_{6c}^{(2)}$, incorporating the spin. Figure 6 schematically illustrates the energy level diagram. Therefore, a band-edge optical absorption at a symmetric location within the Brillouin zone involves a transition from the valence-band state, which has an angular momentum of $J_h = S_h = \frac{1}{2}$ (projection on z,$m_s = \frac{1}{2}$), to the split-off conduction band state, which has $J_e = \frac{1}{2}$ $\left(m_J = \frac{1}{2}\right)$.

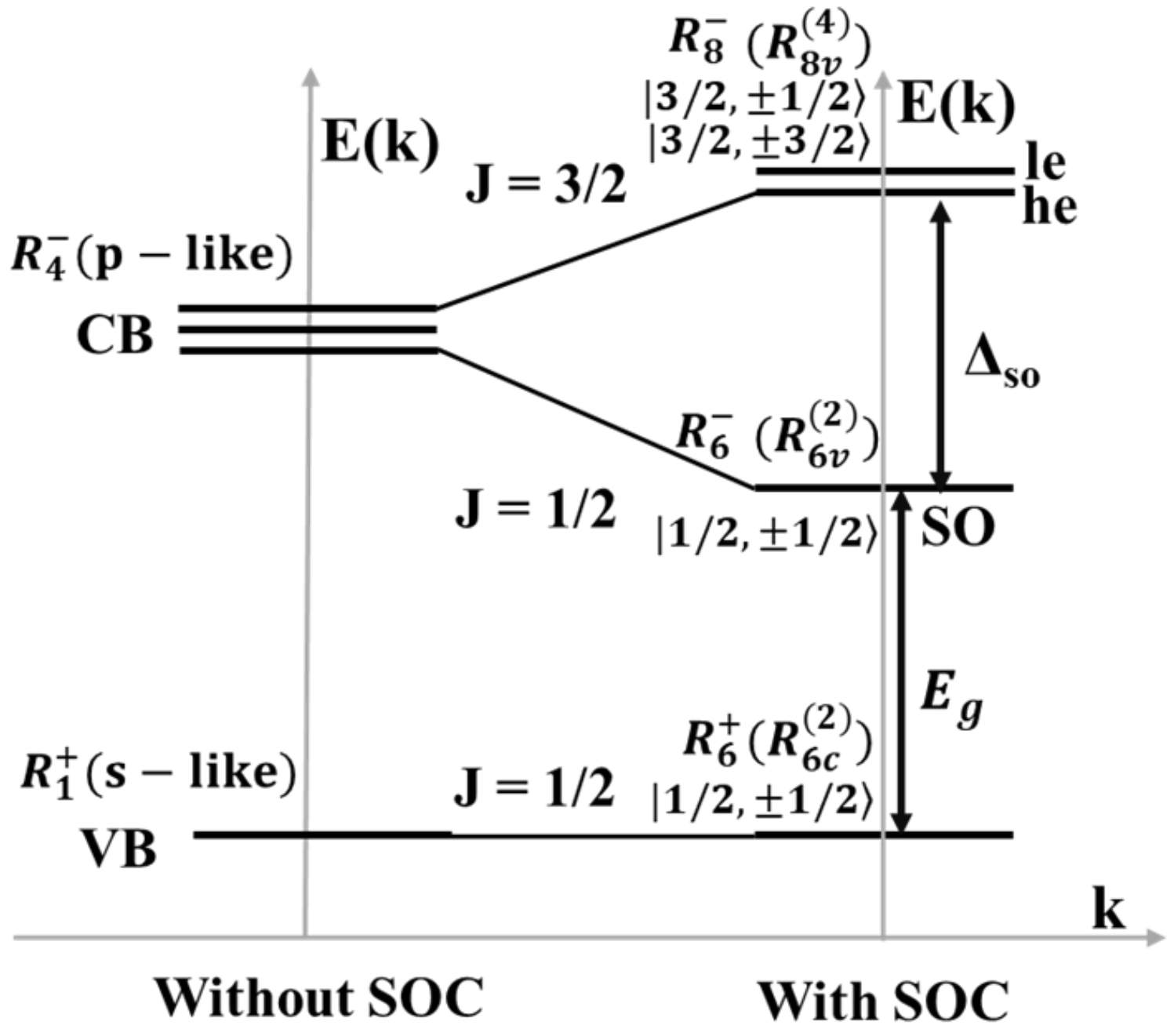

**Figure 6** Schematic band structure of perovskites. The schematic is created using the notation from the ref.[43-44]

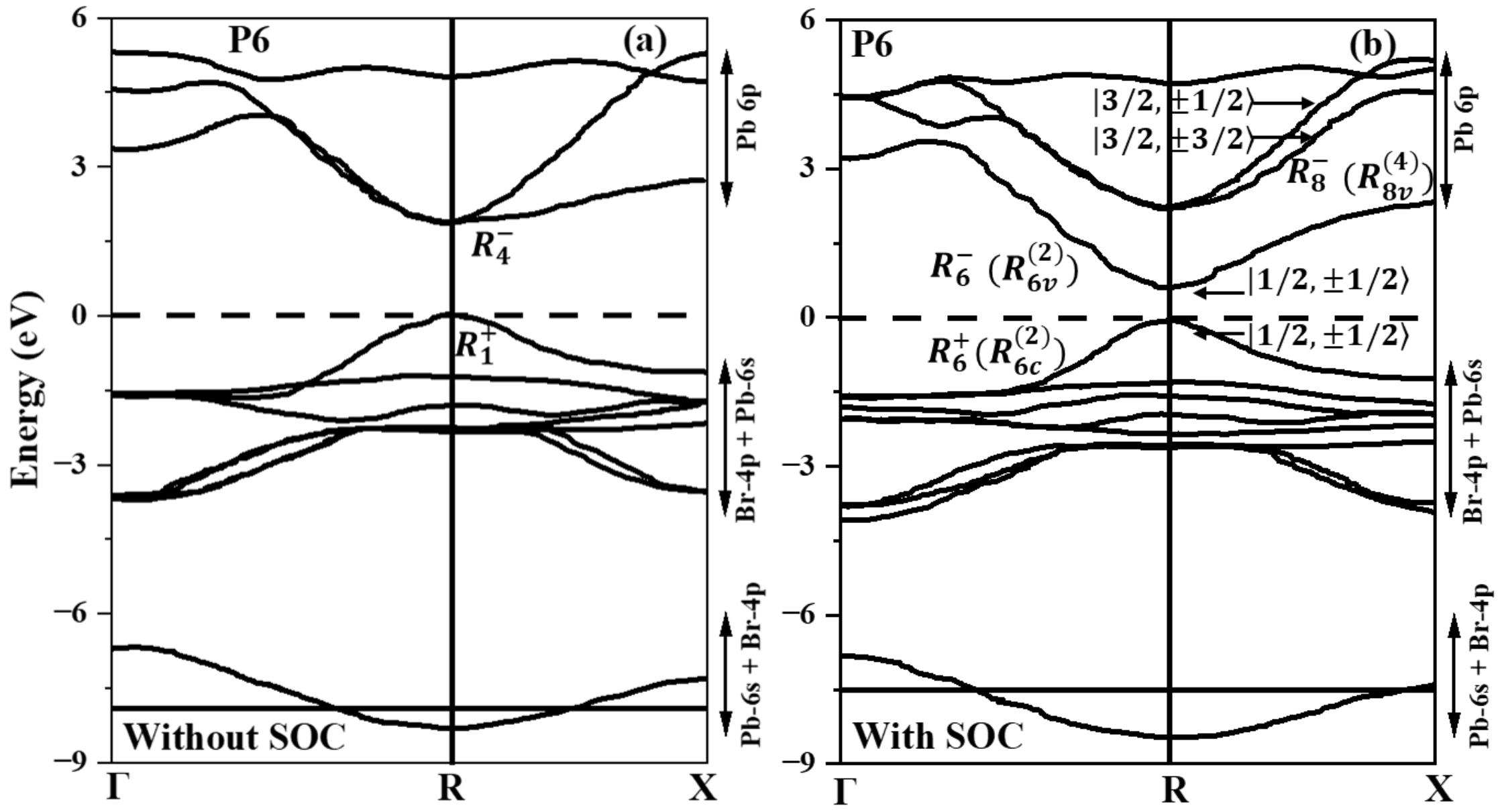

**Figure 7** DFT results for electronic band structure of the $CsPbBr_3$ (P6) (a) without and (b) with spin-orbit coupling (SOC). The electronic band structure is adopted from ref[45] and symmetry notation is applied based on the reference.[43, 45]

---

DFT electronic band structure of P6 is shown in Figure 7a and 7b respectively without and with spin-orbit coupling (SOC) along with the symmetry notation. Since the valence band is made up of the Br 4*p* and Pb 6*s* states, spin-orbit coupling does not have much effect on it. The conduction band, on the other hand, comes from the Pb 6*p* states. The Pb 6*p* orbitals have a strong spin-orbit effect that causes them to split into $J_{3/2}$ and $J_{1/2}$ levels. Akkerman et al.[23] recently provided theoretical insights into the possible electronic transition in lead halide perovskite using effective mass approximation (EMA) calculation, which indicates that the spherical transition amplitude $M_{eh}$ fulfils the following equation 2.

**Equation 2** $$|M_{eh}|^2 = E_P \delta_{l_e l_h} \delta_{m_e m_h} \delta_{n_e n_h}$$

where $E_P$ Kane parameter, the principal quantum numbers of the electron and hole states, respectively, are $n_e$ and $n_h$. According to the aforementioned selection criterion, the only optically permitted transitions in EMA are those denoted by the different spherical angular momenta and principal quantum numbers, such as $1s - 1s, 1p - 1p, 1d - 1d,$ etc. Following the theoretical insight and DFT results reported so far, for the first time, we assign the electronic transitions in all-inorganic perovskite NCs using MCD spectroscopy.

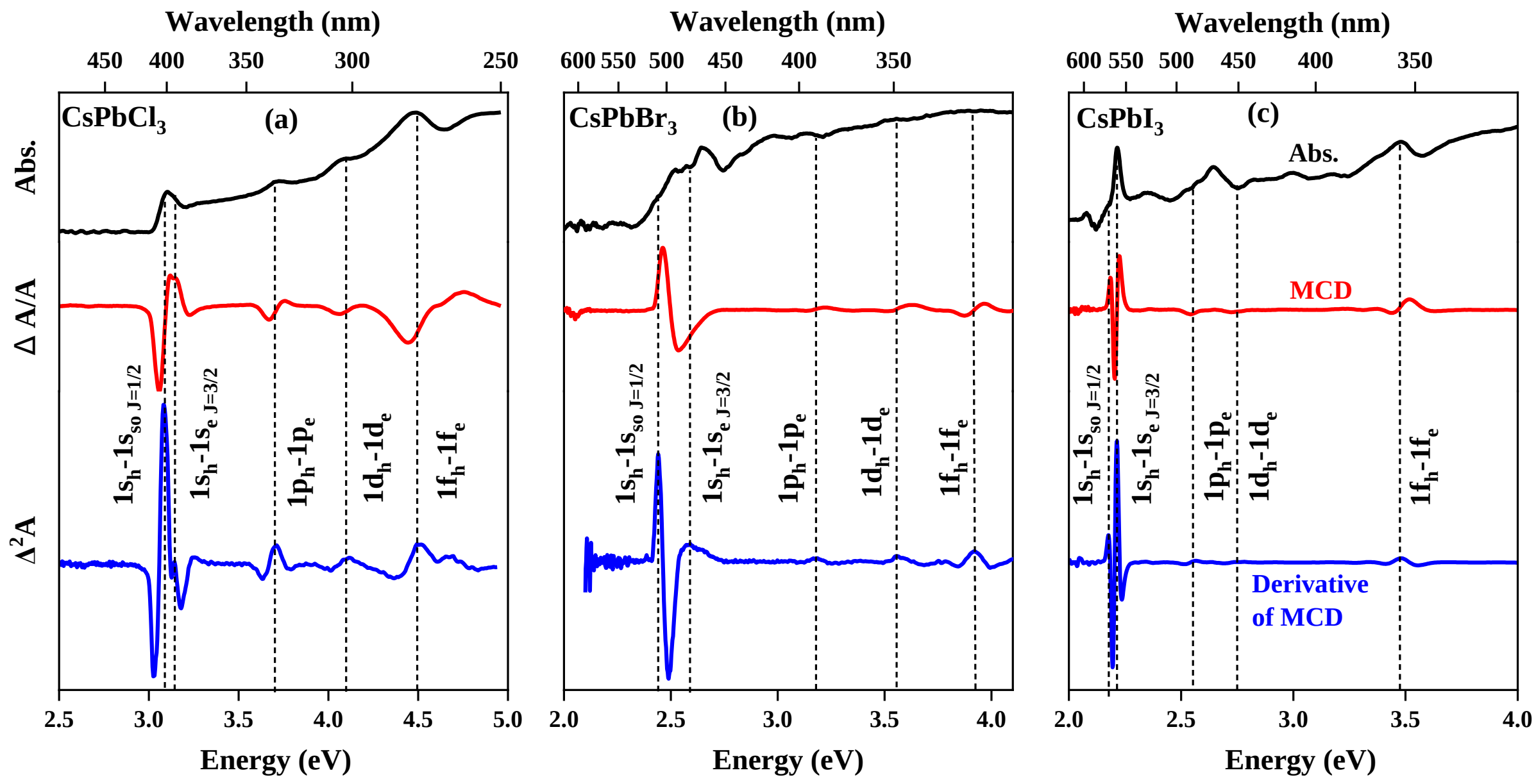

**Figure 8** Probing of different excitonic transitions using MCD spectroscopy. MCD spectra collected at 1.5 K in 5T magnetic field The maxima of the first derivative of the MCD spectra are taken as the transition energy (a)P1- $CsPbCl_3$, (b) P6 - $CsPbBr_3$, and (c) P10 - $CsPbI_3$.

Figure 8 shows the 1.5 K absorption and associated MCD spectra at 5 Tesla of $CsPbCl_3$ (P1), $CsPbBr_3$ (P6), and $CsPbI_3$ (P10) NCs. For each transition in the absorption spectra, we observe a Gaussian or derivative shape signal in the MCD spectra. As previously stated, unlike holes, conduction-band electrons are affected by mixing between sub bands. (light electron, heavy electron, and split-off electron: le, he and so). Conduction band mixing give rise to light electron ($1s_{le_{J=3/2}}$), heavy electron ($1s_{he_{J=\frac{3}{2}}}$) or split- off state ($1s_{so_{J=\frac{1}{2}}}$) .Thus accordingly we assigned the excitonic transitions namely $1s_h - 1s_{so_{J=3/2}}$, $1s_h - 1s_{he_{J=3/2}}$, $1s_h - 1s_{le_{J=1/2}}$, $1p_h - 1p_e$, $1d_h - 1d_e$, $1f_h - 1f_e$ etc. A precise determination of the energy position of each excitonic transition is achieved by analysing the positions of the maxima of the first derivative of the MCD spectra (the blue curve in Figure 8). The assign transitions of $CsPbCl_3$, $CsPbBr_3$ and $CsPbI_3$ NCs are shown in Figure 8 (a-c), each of these transitions is indicated by a dashed line. However, in these cases, we are unable to distinguish between the *h-he* and *h-le* transitions due to their energy levels being very close. We have expanded our work to include mixed halides and have observed that, in the case of these mixed halides, we do observe both *h-he* and *h-le* transitions. Figure 9 shows the absorption, MCD and derivative of MCD spectra of P2, P3 and P4. We can clearly distinguish all the transitions from P3 onwards. However, higher order transitions are not well resolved. This is because all excitonic transitions do not consistently display the same levels of oscillator strength. For instance, transitions such as $1p-1p, 1d-1d, 1f-1f$ display lower oscillator strength, resulting in a weaker signal. However, in case of P1 and P2, the excitonic transitions are well separated and distinguishable. In Figure 10 we have shown the absorption, MCD and derivative of MCD spectra of Br/I mixed halide perovskite NCs. Assigned excitonic transitions are marked with dotted line.

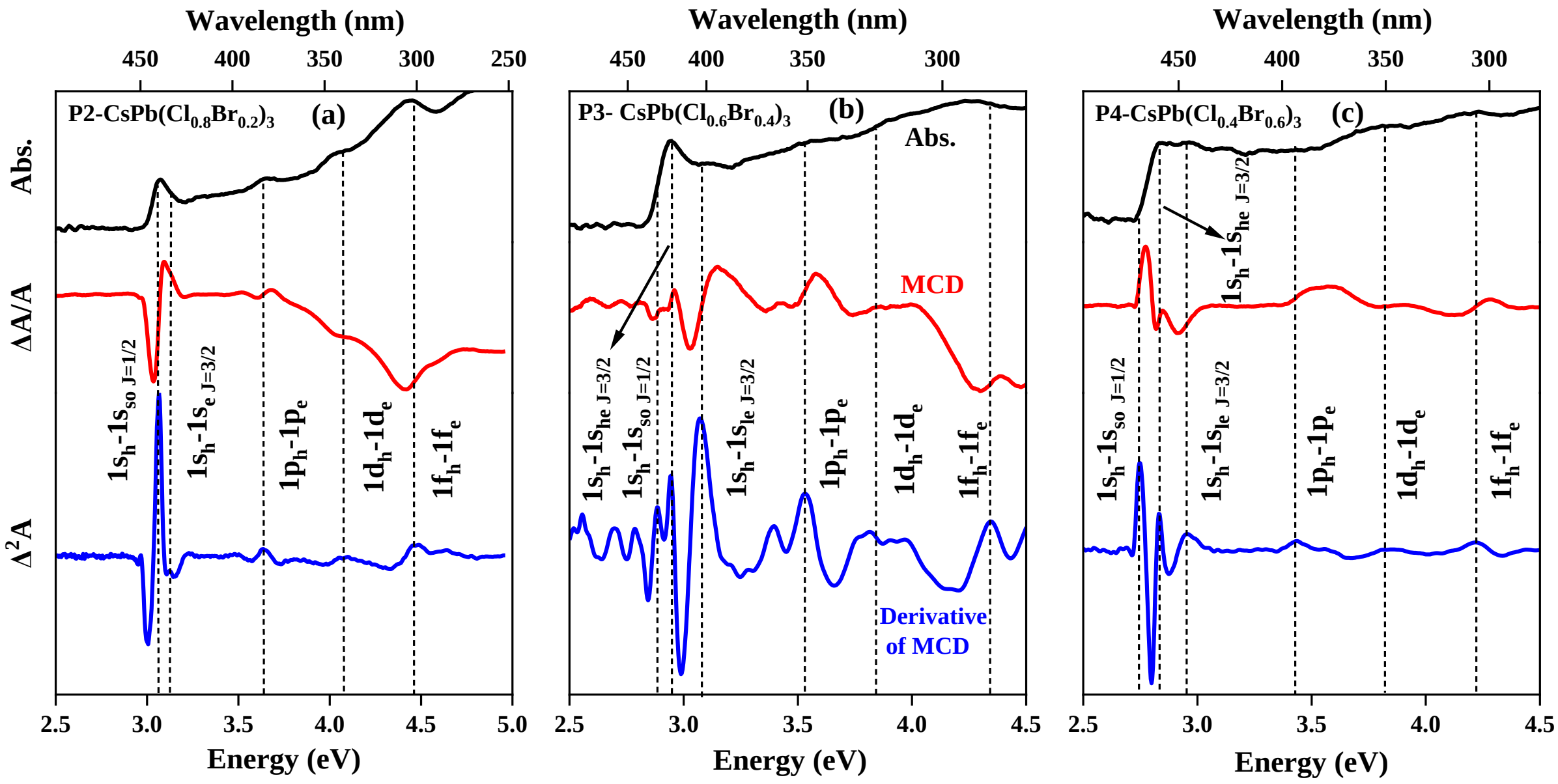

**Figure 9** Probing of different excitonic transitions using MCD spectroscopy in mixed halides. MCD spectra collected at 1.5 K in 5T magnetic field. The maxima of the first derivative of the MCD spectra are taken as the transition energy (a) P2- $CsPb(Cl_{0.8}Br_{0.2})_3$, (b) P3-$CsPb(Cl_{0.6}Br_{0.4})_3$, and (c) P4- $CsPb(Cl_{0.4}Br_{0.6})_3$.

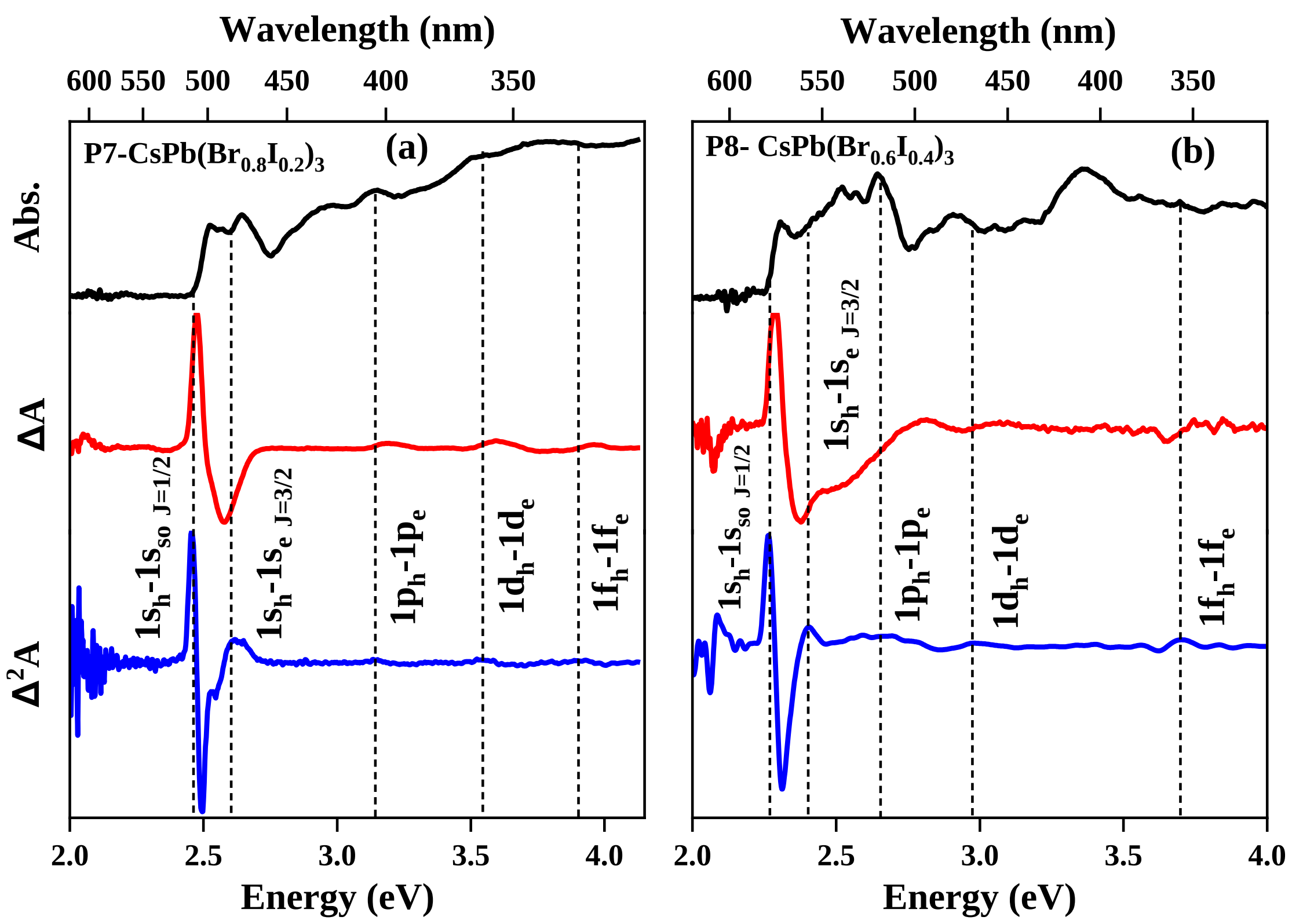

**Figure 10** Probing of different excitonic transitions using MCD spectroscopy in mixed halides. MCD spectra collected at 1.5 K in 5T magnetic field The maxima of the first derivative of the MCD spectra are taken as the transition energy (a) P7- $CsPb(Br_{0.8}I_{0.2})_3$ and (b) P8-$CsPb(Br_{0.6}I_{0.4})_3$.

---

Here we not only probe the lowest energy state but also the higher excited states by doing the magneto-optical measurements of the Zeeman splitting of a large number of perovskites NCs of different compositions using MCD spectroscopy. To probe the nature of these transitions we used variable temperature and variable magnetic field MCD spectroscopy to measure the Zeeman splitting. The excitonic states Zeeman splitting ($\Delta E_z$) calculated using the following equation 3[25]

**Equation 3** $$\Delta E_z = -\left(\frac{\sqrt{2e}}{2}\right)\sigma\frac{\Delta A}{A}$$

Where $\Delta A$ is the maximum excitonic MCD intensity ($\Delta A = A_L - A_R$), and $A$ is the absorbance at the energy of the MCD peak intensity. σ refers to the bandwidth of Gaussians. The conversion $\theta$(mdeg) = 32980$\Delta A$ was used to compute $\Delta A$ values from MCD ellipticity data.[25]

Furthermore, the MCD feature's line shape represents the sign of the excited-state Zeeman splitting. We can anticipate a negative (positive) MCD peak followed by a positive (negative) peak with increasing energy for a positive (negative) Zeeman splitting. In figure 11 we have shown the Zeeman splitting of the $1s_h - 1s_{so_{J=3/2}}$ transition with varying the magnetic field and temperature for P1, P6 and P10. The result indicates that in P1, $\Delta E_z$ is independent of temperature and increases linearly with the magnetic field. However, in P6 and P10, as depicted in Figure 11(b-c), the sign of $\Delta E_z$ is opposite (negative) and demonstrates temperature-dependent behaviour. The change in polarization of the excited state due to compositional variation results in a change in sign and temperature dependency, which arises from the contribution induced by the heavier halides (Br, I) towards the split off state. Thus, the composition of the perovskite NCs plays a significant role in determining the nature of MCD signal. To verify that we conducted additional investigations into the nature of the $1s_h - 1s_{so_{J=1/2}}$ transition in different mixed halide NCs, as illustrated in Figure 12. The sign inversion of the $\Delta E_z$ is observed in these NCs with changing temperature. Below 5K $\Delta E_z$ is negative and then above 5K it exhibits the positive sign. On the other hand, $1s_h - 1s_{e_{J=3/2}}$ transitions (2nd excitonic transitions) in P3, P4, P7, and P8 NCs that do not show sign change in $\Delta E_z$ when the temperature changes as shown in Figure 13. The $\Delta E_z$ is dominated at low temperature and there is a small temperature dependence because of the mixing of heavy and light electronic states with the split-off states.

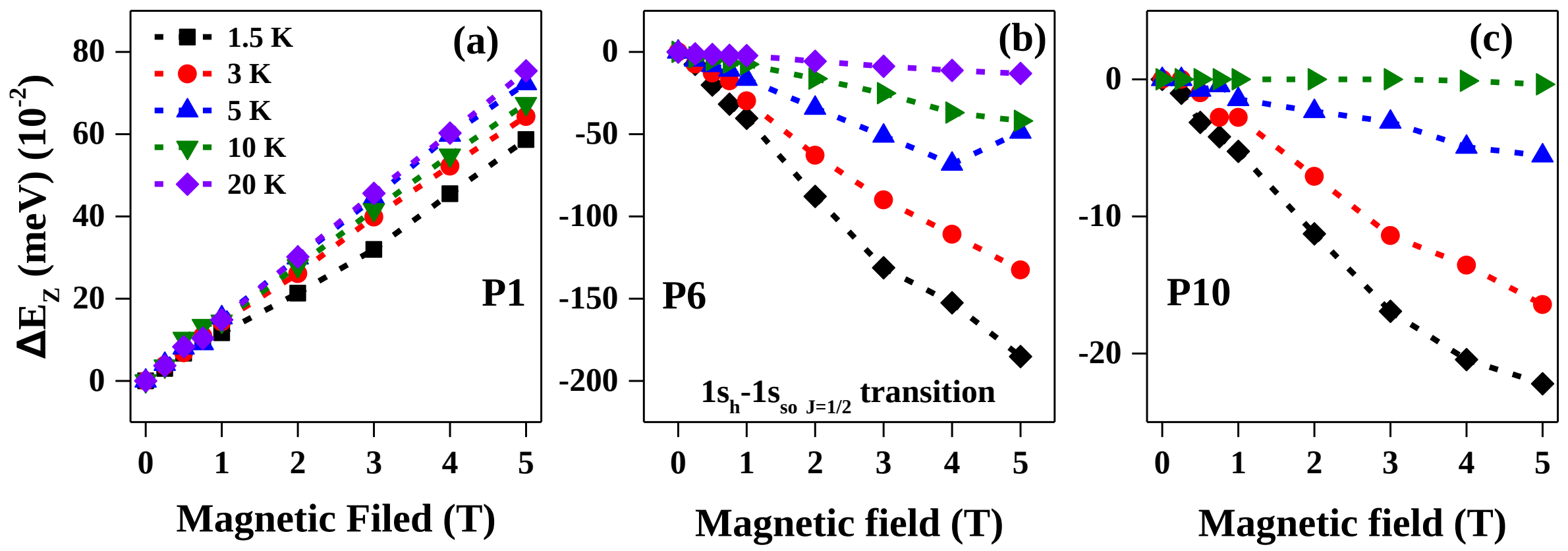

**Figure 11** Variation of Zeeman energy with magnetic field and temperature for the $1s_h - 1s_{so_{J=1/2}}$ transition for (a) $CsPbCl_3$ (P1), (b) $CsPbBr_3$ (P6), and (c) $CsPbI_3$ (P10).

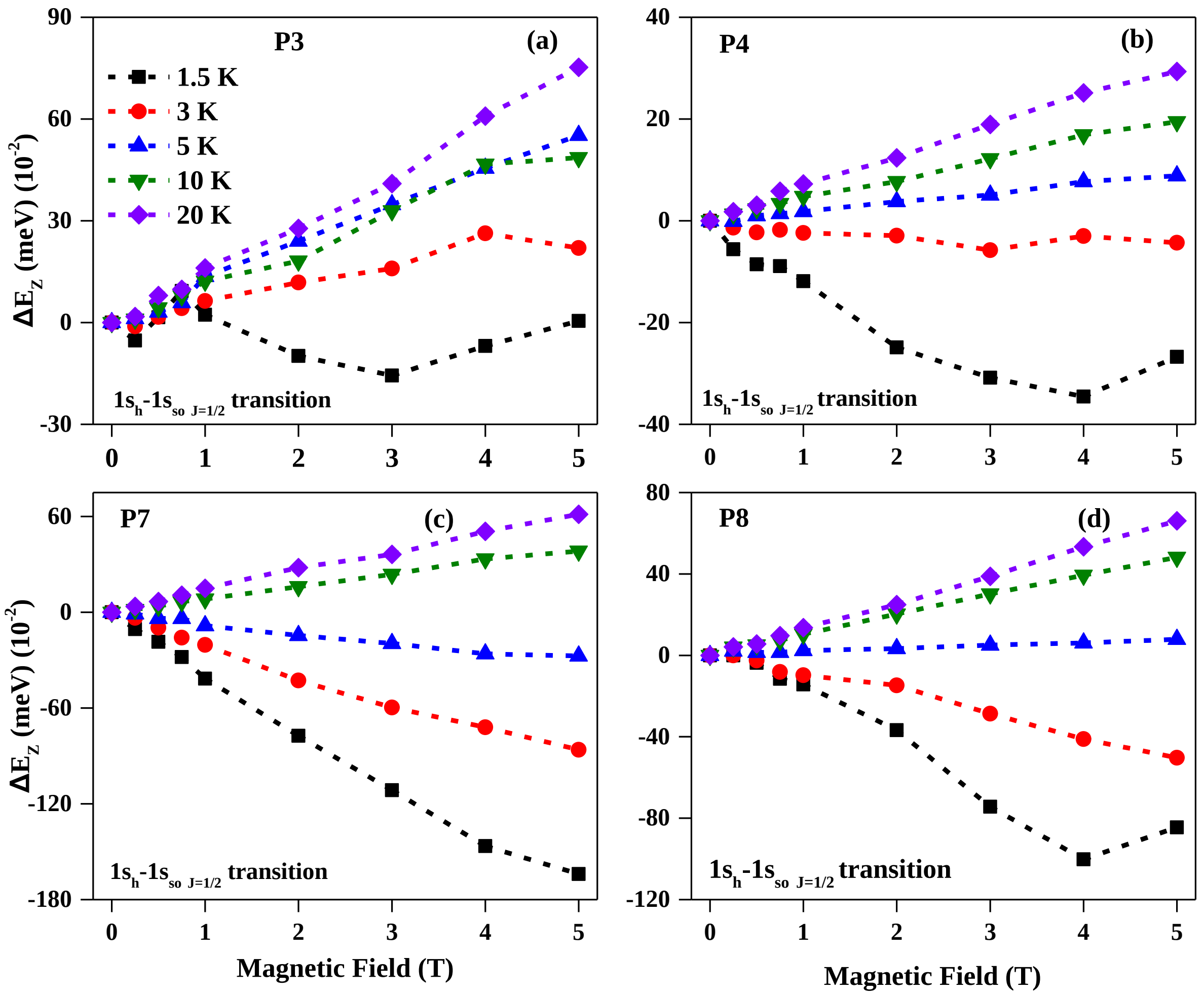

**Figure 12** Variation of Zeeman energy with magnetic with and temperature for the $1s_h - 1s_{SO_{J=1/2}}$ transition for (a) P3- $CsPb(Cl_{0.6}Br_{0.4})_3$, (b)P4 - $CsPb(Cl_{0.4}Br_{0.6})_3$, (c) P7-$CsPb(Br_{0.8}I_{0.2})_3$, and (d) P8 - $CsPb(Br_{0.6}I_{0.4})_3$.

---

In Figure 13, we show a comparison of the $1s_h - 1s_e$ transitions energy across all of the samples of perovskite, which originated by splitting in the conduction band. We are able to separate the I. $1s_h - 1s_{he_{J=3/2}}$, II. $1s_h - 1s_{le_{J=1/2}}$, and III. $1s_h - 1s_{so_{J=1/2}}$transitions through the utilization of MCD spectroscopy. In P3 and P4, we were able to witness all three transitions. In the cases of P1-P2, P6 to P10, we are unable to differentiate between I and II transition due to the overlapping energy band.

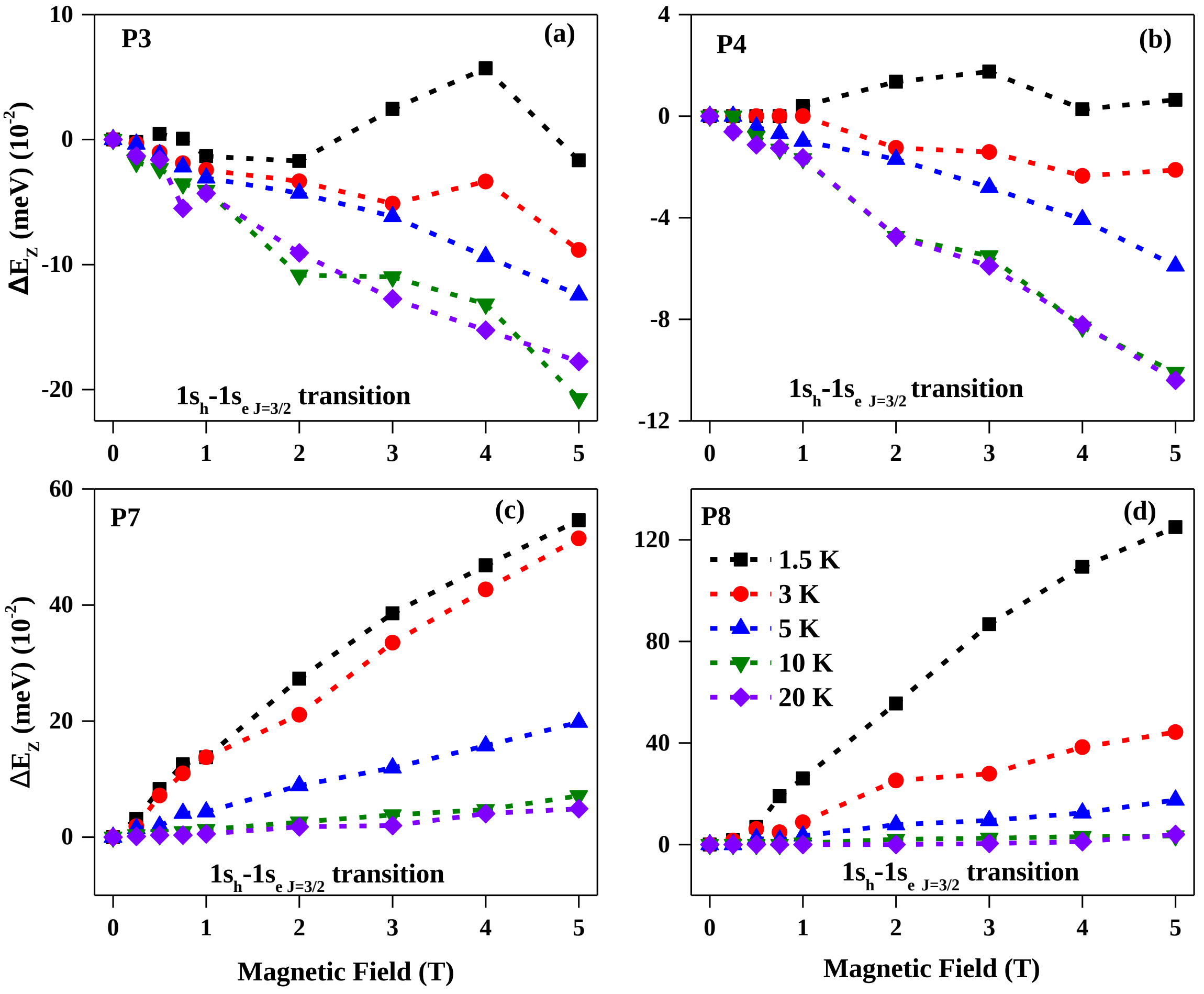

**Figure 13** Variation of Zeeman energy with magnetic with and temperature for the $1s_h - 1s_{e_{J=3/2}}$ transition for (a)P3- $CsPb(Cl_{0.6}Br_{0.4})_3$, (b) P4- $CsPb(Cl_{0.4}Br_{0.6})_3$, (c) P7- $CsPb(Br_{0.8}I_{0.2})_3$, and (d) $CsPb(Br_{0.6}I_{0.4})_3$.

---

To explain the origin of Zeeman splitting in perovskite NCs and their temperature-dependent behaviour, we have illustrated the process in schematic 1. In the presence of a magnetic field, the valence and conduction bands split into two Zeeman components. According to the selection rule $\Delta m_J = \pm 1$ for MCD in the Faraday geometry, two transitions are permitted. The Zeeman energy can be described using the equation 4.

**Equation 4** $$\Delta E_Z = g\mu_B B$$

(Where $g$ = exciton g-factor, $\mu_B$= Bohr magneton, $B$= applied magnetic field). It is independent of temperature. The $\Delta E_Z$ of P1 for the $1s_h - 1s_{so_{J=1/2}}$ transitions can be explained using the

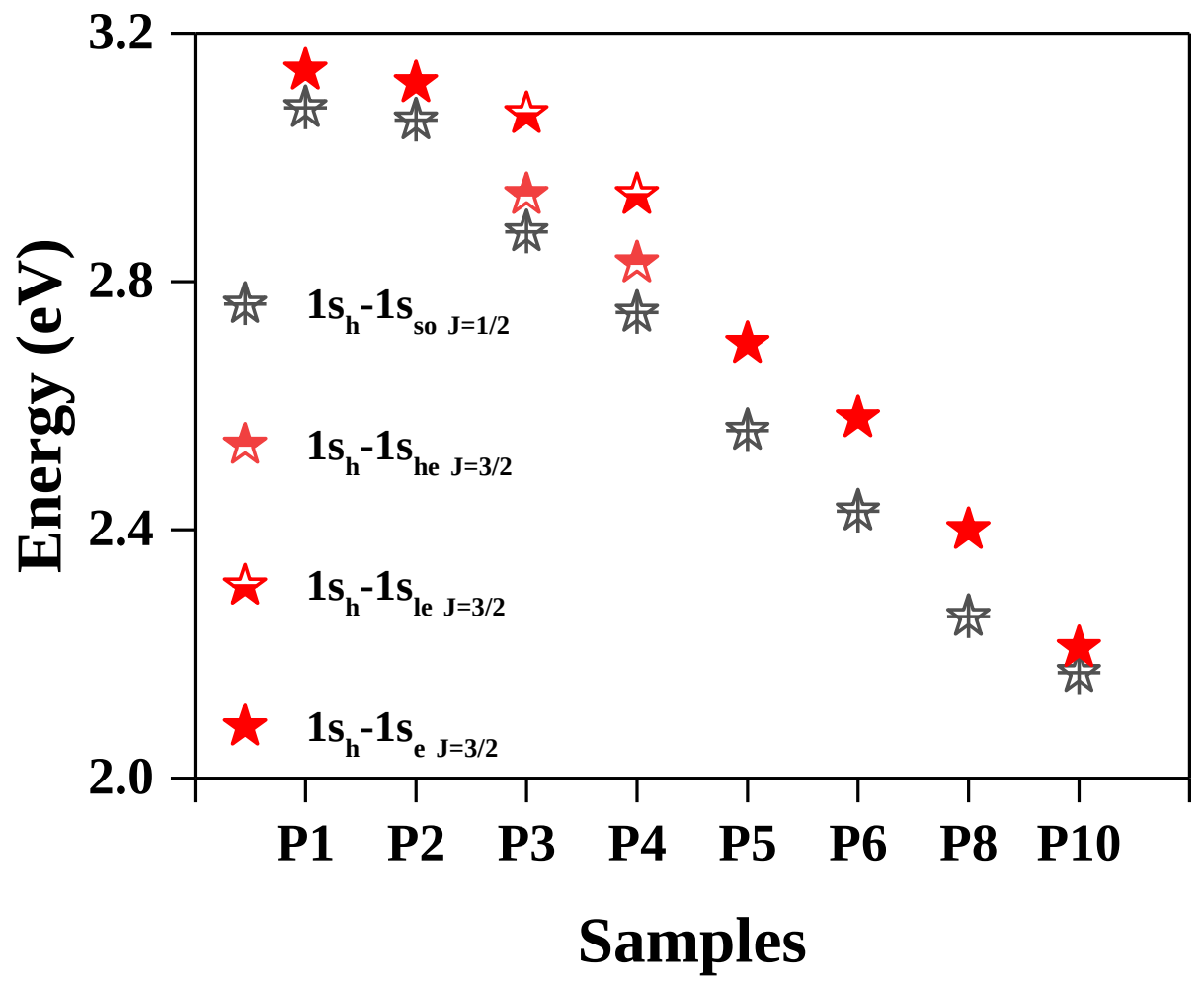

**Figure 14** Comparison of $1s_h - 1s_e$ transitions energy over all the perovskite NCs

---

equation 4. As in $CsPbCl_3$ NCs the effect of halide contribution towards the split off state is negligible and the origin of Zeeman splitting is purely magnetic field induced and it follow the linear relationship with the increase in magnetic field. However, with the introduction of bromide into the system we observe the split-off state induced $1s_h - 1s_{so_{J=1/2}}$ transitions become temperature dependent. The Zeeman energy of this transition can be illustrated using the equation 5.

**Equation 5** $$\Delta E_Z = g\mu_B B + \Delta_{so}$$

Now there will be two contributions on exciton Zeeman splitting – i) intrinsic splitting ($g\mu_B B$) and ii) split off state induced contribution ($\Delta_{so}$). Among these $\Delta_{so}$ is temperature dependent. Split off state give negative Zeeman energy. Thus, depending on temperature one of the contributions is dominant. From the data shown in Figure 12 it is evident that below 5K,

**Schematic 1** Exciton Zeeman splitting mechanism of lead halide perovskite NCs.

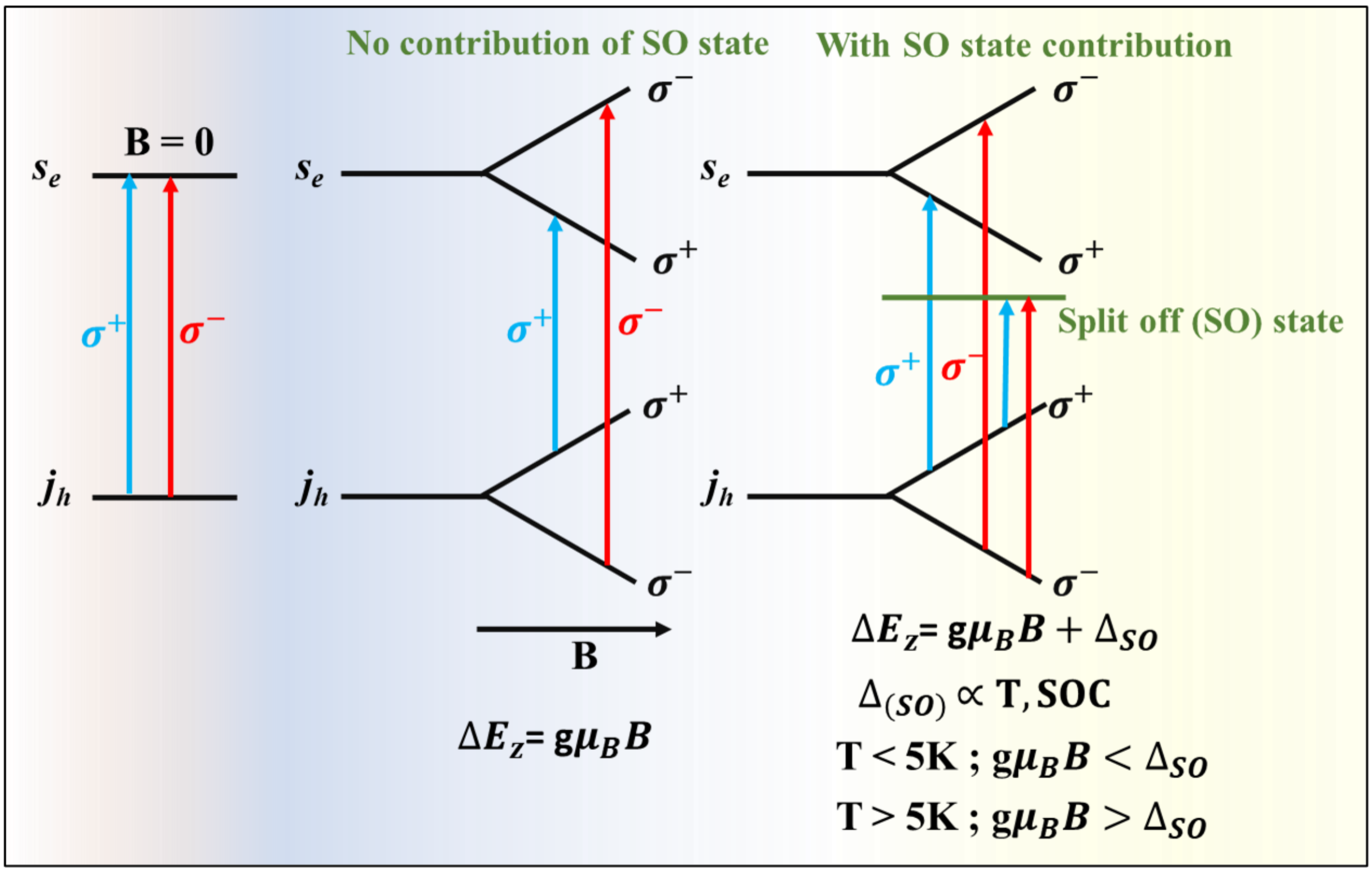

$g\mu_B B < \Delta_{so}$ thus $\Delta E_Z$ become negative and above 5K, $g\mu_B B > \Delta_{so}$ thus $\Delta E_Z$ become positive. Overall, our data clearly depict that perovskite NCs' behaviour is significantly different compared to II-VI semiconductor NCs and heavier halide induced contribution towards spin-off state and spin-orbit coupling play a major role in describing the nature of exciting transitions.

## Conclusions

Colloidal NCs of $CsPbX_3$ were synthesized using the hot-injection method. Spectroscopic investigations into the electronic structure have provided an experimental description of the magnetic-exciton interaction in these interesting materials. Using absorption and MCD spectroscopy we successfully assigned up to six excitonic transitions. To probe the nature of these transitions we studied the first two excitonic transitions (as other transitions are not well resolved) using variable temperature and variable magnetic field MCD spectroscopy and revealed that split-off state induced Zeeman splitting is temperature dependent and change its polarity depending on the temperature. The findings show that the Zeeman splitting is influenced by both intrinsic and split-off-state-induced factors. The results show that the exciton spin polarisation can be tuned and inverted with the change in temperature and magnetic field, which could be important for future spin-photonic information processing technologies that rely on spin manipulation within QDs.

## Author Contributions

The project designs were formulated by P.M. and R.V. P.M. was responsible for sample synthesis and conducted all measurements. Both P.M. and R.V. reviewed the data analysis and interpretation. The first draft was written by P.M., with subsequent manuscript revisions contributed to by both researchers.

## Notes

The authors declare no competing financial interest.

## Acknowledgement

The authors express their gratitude to SERB (CRG/2018/000651), SERB-POWER Fellowship (SPF/2021/000110), ICMS, JNCASR, and the Department of Science and Technology, Government of India, particularly Prof. C. N. R. Rao, for their generous financial support. P.M. extends thanks to CSIR for providing a research fellowship. The authors also recognize the Technical Research Centre Microscopy lab at JNCASR Bangalore for their assistance with TEM imaging.